\documentclass [prb,twocolumn,superscriptaddress,showkeys,preprintnumbers]{revtex4}
\usepackage {bm}
\usepackage {graphicx}

\usepackage {amssymb}
\usepackage {amsmath}

\begin{document}

\title{Phase Aberrations in Diffraction Microscopy}

\author{S.~Marchesini} \email{smarchesini@llnl.gov}
\author{H.~N.~Chapman}
\affiliation{University of California, Lawrence Livermore National
  Laboratory, 7000 East Ave., Livermore, CA 94550, USA}
\affiliation{Center for Biophotonics Science and Technology, UC Davis, 2700 Stockton Blvd., Ste 1400, Sacramento CA, USA}
\author{A.~Barty}
\affiliation{University of California, Lawrence Livermore National
  Laboratory, 7000 East Ave., Livermore, CA 94550, USA}
\author{C.~Cui}
\author{M.~R.~Howells}
\affiliation{Advanced Light Source, Lawrence Berkeley National Laboratory,
  1 Cyclotron Road, Berkeley, CA 94720, USA}
\author{J.~C.~H.~Spence}
\author{U.~Weierstall}
\affiliation{Department of Physics and Astronomy, Arizona State University,
  Tempe, AZ 85287-1504, USA}
\author{A.~M.~Minor} 
\affiliation{National Center for Electron Microscopy, Lawrence Berkeley National Laboratory,   1 Cyclotron Rd, Berkeley, CA 94720, USA}

\begin{abstract}
In coherent X-ray diffraction microscopy the diffraction pattern generated 
by a sample illuminated with coherent x-rays is recorded, and a computer 
algorithm recovers the unmeasured phases to synthesize an image. By avoiding 
the use of a lens the resolution is limited, in principle, only by the 
largest scattering angles recorded. However, the imaging task is shifted 
from the experiment to the computer, and the algorithm's ability to recover 
meaningful images in the presence of noise and limited prior knowledge may 
produce aberrations in the reconstructed image. We analyze the low order 
aberrations produced by our phase retrieval algorithms. We present two 
methods to improve the accuracy and stability of reconstructions. 
\end{abstract}


\preprint{UCRL-PROC-215873}
\maketitle

\section{Introduction}

A new imaging technique has emerged in recent years that can overcome many 
limitations of light, electron, and X-ray microscopy. Coherent X-ray 
Diffraction Microscopy (CXDM) \cite{Miao:1999} promises to enable the study of thick 
objects at high resolution. In this technique one records the 3D diffraction 
pattern generated by a sample illuminated with coherent x-rays, and as in 
x-ray crystallography a computer recovers the unmeasured phases. This is 
done by alternately applying constraints such as the measured intensity in 
reciprocal space and the object support---the region where the object is 
assumed to be different from 0---in real space. This corresponds to defining 
the envelope of a molecule in crystallography. In our implementation the 
support is periodically updated based on the current object estimate \cite{Marchesini:2003}.

By avoiding the use of a lens, the experimental requirements are greatly 
reduced, and the resolution becomes limited only by the radiation damage 
\cite{Marchesini:2004,Howells:2004}. However the imaging task is shifted from the experiment to the 
computer, and the technique may be limited by our understanding of the phase 
recovery process as well as the algorithm's ability to recover meaningful 
images in the presence of noise and limited prior knowledge.

Recently we have presented experimental results of high-resolution 3D X-ray 
diffraction imaging of a well-characterized test object to demonstrate the 
practical application of these advances \cite{Chapman:2005,Marchesini:IT23}. Here we extend the 
analyis of image reconstruction and determine low-order phase errors 
(essentially image aberrations) that can occur when reconstructing general 
complex-valued images. We present two methods to improve the accuracy and 
stability of reconstructions.

\section{Coherent X-Ray Diffraction}
\begin{figure}[htbp]
\centerline{\includegraphics[width=0.4\textwidth]{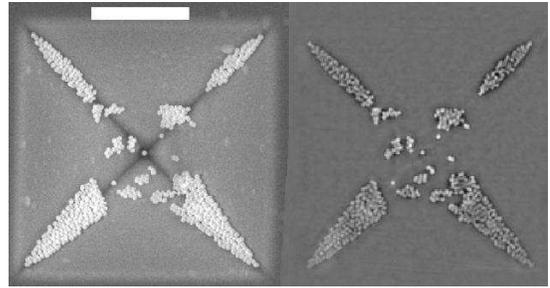}}
\caption{(a) SEM image of the 3D test object.  Scalebar is 1
micron. (b) Infinite depth-of-focus image reconstructed from a central
section of the 3D coherent X-ray diffraction data.}  \label{fig1}
\end{figure}

Three-dimensional coherent X-ray diffraction data were collected at the 
Advanced Light Source \cite{Beetz:2005,Howells:2002} from a test object that consisted of 50-nm 
diameter gold spheres located on a 2.5-$\mu $m-wide silicon nitride pyramid 
\cite{Chapman:2005} (Fig. \ref{fig1}a). A bare CCD lo,cated in the far field recorded the 
diffraction patterns with a pixel sampling that was more than 4 times the 
Shannon sampling rate for the (phased) complex amplitudes. Diffraction 
patterns were collected for many sample orientations over an angular range 
of 129\r{ }. These were interpolated onto a 3D grid. We reconstructed a full 
3D image by performing phase retrieval on the entire 3D diffraction dataset 
(i.e. the iterations involved three-dimensional FFTs). The resulting volume 
image reveals the structure of the object in all three dimensions and can be 
visualized in many ways including projections through the data, slices 
(tomographs), or isosurface rendering of the data.

In addition to 3D images, we perform much analysis and algorithm development 
on 2D datasets. For the work in this paper we choose central plane sections 
extracted from the 3D diffraction pattern. By the Fourier projection 
theorem, the image formed from a central section is an infinite 
depth-of-focus projection image (Fig. \ref{fig1}b). We carry out \textit{ab initio} image 
reconstructions using the Relaxed Averaged Alternating Reflections (RAAR) 
algorithm \cite{Luke:2005} with the ``Shrinkwrap'' dynamic support constraint 
\cite{Marchesini:2003}. Details of the algorithm parameters used are given in Chapman \cite{Chapman:2005}.

\section{Resolution Analysis}

The phase retrieval process recovers the diffraction phases with limited 
accuracy, due to factors including SNR of the diffraction amplitudes, 
missing data, the inconsistency of constraints, and systematic errors in the 
data (such as errors in interpolation). These errors in phase reduce the 
resolution of the synthesized image. With a complex image a loose support 
constraint will lead to unconstrained low-order aberrations. As is well 
known an object could be shifted by a few pixels each time we reconstruct, 
which is equivalent to a varying linear phase ramp in reciprocal space. In 
addition to this shift low order phase variations, such as defocus and 
astigmatism can also be unconstrained if the aberrated object fits inside 
the support. One way to quantify the effect of these phase variations is to 
determine the variation in retrieved phases as a function of resolution 
\cite{Shapiro:2005}
$^{10)}$. Given a reconstructed image $g(x)$ obtained by phase retrieval 
starting from random phases, and its Fourier transform $G = \left| G 
\right|\exp \left\{ {i\varphi (q)} \right\}$, we define the phase retrieval 
transfer function by

\begin{equation}
\label{eq1}
PRTF\left( {\rm {\bf q}} \right) = \left| {\left\langle {\exp \left\{ 
{i\varphi ({\rm {\bf q}})} \right\}} \right\rangle } \right| = \left| 
{\left\langle {\frac{G\left( {\rm {\bf q}} \right)}{\left| {G\left( {\rm 
{\bf q}} \right)} \right|}} \right\rangle } \right|
\end{equation}

\noindent
with $\left\langle G \right\rangle $ the average over the complex 
diffraction amplitudes of many reconstructed images starting from random 
phases. Where the phases are random and completely uncorrelated, the average 
will approach zero. Thus, the ratio is effectively a transfer function for 
the phase retrieval process, and the average image (the Fourier tranform 
of $\left\langle G \right\rangle )$ is the best estimate of the image: 
spatial frequencies are weighted by the confidence in which their phases are 
known.

\begin{figure}[htbp]
\centerline{\includegraphics[width=0.25\textwidth]{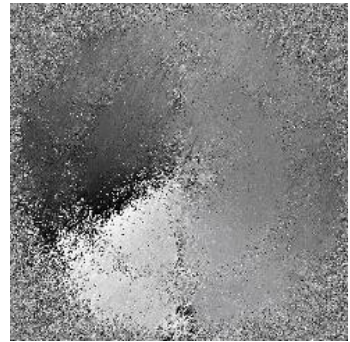}}
\caption{Phase difference between two reconstructions in reciprocal
space, showing a phase vortex between two solutions in the far field.
The center of the vortex is at q = 0, and the half-width of the phase
map is q = 0.048 nm$^{-1}$.}\label{fig2} 
\end{figure}

In our case when reconstructing complex 2D images, with low frequencies 
missing due to the beamstop, we have observed that phase retrieval from 
independent random starts may differ by a phase vortex (right or left 
handed), centered at the zero spatial frequency (Fig. \ref{fig2}). We find that we 
can improve the estimate of the image by separating out the vortex modes 
\cite{Chapman:2005}. 
These phase vortices are due to stagnation of the phase retrieval 
process. Other phase vortices can appear near local minima of the measured 
intensities, and our method of separating solutions will fail to detect 
vortices not centered near the beamstop. In order to remove these vortex 
aberrations we modified the reconstruction algorithm as follows: (i) Average 
$n$ independent reconstructions which will likely average out the phase vortex 
modes but will also smooth the resulting image, reducing the resolution. 
(ii) Refine this averaged image by inputting it to the RAAR \cite{Luke:2005}
algorithm and carrying out 200 iterations. Using this ``averaged RAAR'' 
algorithm we reduced the probability of recovering an image with phase 
vortex mode from 40{\%} to 15{\%}, resulting in an improvement of the PRTF 
by almost a factor of two.

\begin{figure}[htbp]
\centerline{\includegraphics[width=0.45\textwidth]{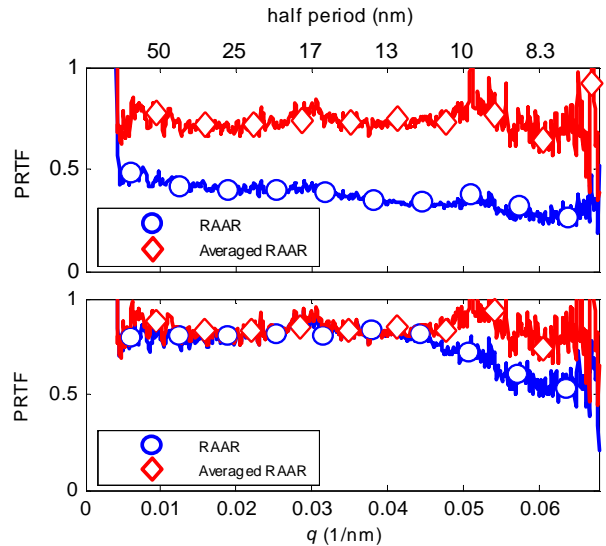}}
\caption{Top: Phase retrieval transfer function of the two
algorithms. PRTF=1 represents stable phases. The averaged RAAR
algorithm significantly improves the stability of the
reconstructions. Bottom: PRTF after removing the vortices centered at
q = 0. The averaged RAAR shows marked improvements at high
frequencies.}\label{fig3} 
\end{figure} 

We compute the final image, and the PRTF, by averaging 1000 such 
reconstructions (Fig. \ref{fig3}). Before averaging many images we make sure that 
they are not shifted with respect to one another by finding the linear phase 
ramp that minimizes the difference between their Fourier transforms. 
Fluctuations of the linear phase term indicate fluctuations in positions. 
Fluctuations in higher order polynomial phase terms indicate that phase 
aberrations are present in the reconstructions.

To quantify the instabilities of these low order phase modes, we find the 
low order phase modes (focus, astigmatism, coma, up to a polynomial of order 
$n_{p})$ that minimize the difference between each new reconstruction 
$G_{n}$ and the first recovered image $G_{0}$. This is done by minimizing

\begin{equation}
\label{eq2}
\chi = \sum\limits_{{\rm {\bf q}}} {\left| {G_{0} \left( {\rm {\bf q}} 
\right) - G_{n} \left( {\rm {\bf q}} \right)\exp \left\{ {ip({\rm {\bf q}})} 
\right\}} \right|^2} 
\end{equation}

\noindent
with the 2D polynomial defined by coefficients $p_{i,j }$ as

\begin{equation}
\label{eq3}
p\left( {\rm {\bf q}} \right) = \sum\limits_{i,j = 0}^{i + j \leqslant n_{p} 
} {p_{i,j} \,\bar {q}_{x}^{i} \,\bar {q}_{y}^{j} } 
\end{equation}

\noindent
with $\bar {q}_{x,y} = {q_{x,y} } \mathord{\left/ {\vphantom {{q_{x,y} } 
{2\max \left( {q_{x,y} } \right)}}} \right. \kern-\nulldelimiterspace} 
{2\max \left( {q_{x,y} } \right)}$. The linear terms representing shifts in 
real space are found using the method described by Fienup \cite{Fienup:1997},
 while 
higher order terms are obtained by fitting the phase difference, $\arg 
(G_{0}^{\dag } G_{n} )$, to the higher order 2D polynomial terms and 
iterating until the correction is less than 1\r{ }. The fluctuations of the 
second order polynomial coefficients are obtained by calculating their 
standard deviation among 1000 reconstructions, and we find that

\begin{equation}
\label{eq4}
std\left( {{\begin{array}{*{20}c}
 {p_{{0,0}} } \hfill & {p_{{1,0}} } \hfill & 
{p_{{2,0}} } \hfill \\
 {p_{{0,1}} } \hfill & {p_{{1,1}} } \hfill & 
{p_{{2,1}} } \hfill \\
 {p_{{0,2}} } \hfill & {p_{{1,2}} } \hfill & 
{p_{{2,2}} } \hfill \\
\end{array} }} \right) = \left( {{\begin{array}{*{20}c}
 {{2.26}} \hfill & {{0.31}} \hfill & {{0.15}} \hfill \\
 {{0.53}} \hfill & {{0.61}} \hfill & {0} \hfill \\
 {{0.14}} \hfill & {0} \hfill & {0} \hfill \\
\end{array} }} \right)
\end{equation}

The linear terms ($p_{1,0}$ $p_{0,1})$ represent a shift of ${\left[ 
{{0.31, 0.53}} \right]} \mathord{\left/ {\vphantom {{\left[ 
{{0.31, 0.53}} \right]} {2\pi = \left[ {{0.049,0.085}} \right]}}} 
\right. \kern-\nulldelimiterspace} {2\pi = \left[ {{0.049,0.085}} 
\right]}$ pixels in real space corresponding to 0.5 and 0.8 nm shifts in $x$ 
and $y$. The degree of defocus phase variation depends on $(p_{2,0} + p_{0,2} ) 
/ 2$, and the real-space defocus variation is given by:

\begin{equation}
\label{eq5}
\delta {\kern 1pt} z = \frac{\lambda }{4\pi NA^2}std\left\{ {p_{2,0} + 
p_{0,2} } \right\}
\end{equation}

We have NA=0.084 and $\lambda $=1.65 nm, giving $\delta {\kern 1pt} z = 
11.3$ nm. Note that this defocus variation represents an instability of the 
phase retrieval process and does not correspond to an optical effect of 
focusing through a thick object. In this case all voxels of the 3D images or 
pixels of the 2D projection images are equally aberrated by this effect. 

\begin{figure}[htbp]
\centerline{\includegraphics[width=0.45\textwidth]{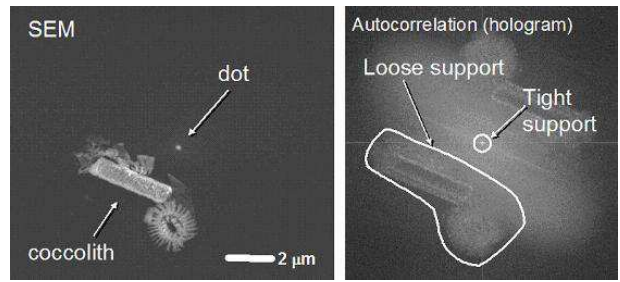}}
\caption{Demonstration of reference-enhanced diffraction imaging.  A
Pt dot was deposited near the sample (a coccolith) with a focused ion
beam instrument.  The Fourier transform of the diffraction intensities
(right) can be used to determine the support which can constrain the
low-order phase aberrations.}
\label{fig4} 
\end{figure}

An additional method to further reduce these instabilities is to use a small 
reference point near the specimen. During the retrieval process the image of 
the reference point is forced to be small with a tight support. This 
constrains the aberrations at this image point, and hence at all image 
points. The reference point has the additional advantage of providing a 
hologram of the specimen (Fig. \ref{fig4}) which can be used to provide the object 
support, or even the desired image.

To quantify our ability to recover unmeasured intensities (for example 
behind the beamstop) we use the normalized standard deviation

\begin{equation}
\label{eq6}
\sigma ^2\left( {\rm {\bf q}} \right) = {\left\langle {\left| {G\left( {\rm 
{\bf q}} \right) - \bar {G}\left( {\rm {\bf q}} \right)} \right|^2} 
\right\rangle } \mathord{\left/ {\vphantom {{\left\langle {\left| {G\left( 
{\rm {\bf q}} \right) - \bar {G}\left( {\rm {\bf q}} \right)} \right|^2} 
\right\rangle } {\left| {\left\langle {G\left( {\rm {\bf q}} \right)} 
\right\rangle } \right|^2}}} \right. \kern-\nulldelimiterspace} {\left| 
{\left\langle {G\left( {\rm {\bf q}} \right)} \right\rangle } \right|^2}.
\end{equation}

We define a transfer function, based on $\sigma ^2$ as:

\begin{equation}
\label{eq7}
TF\left( {\rm {\bf q}} \right) = \frac{1}{\sqrt {1 + \sigma ^2\left( {\rm 
{\bf q}} \right)} } = \frac{\left| {\left\langle {G\left( {\rm {\bf q}} 
\right)} \right\rangle } \right|}{\sqrt {\left\langle {\left| {G\left( {\rm 
{\bf q}} \right)} \right|^2} \right\rangle } }
\end{equation}

\noindent
which has the desired properties that transfer function is unity for $\sigma 
^2 = 0$ and zero for $\sigma ^2 = \infty$. Eqn. (\ref{eq7}) reduces to the PRTF in 
the regions of \textbf{q} where $\vert G\vert $ is measured. 

\begin{figure}[htbp]
\centerline{\includegraphics[width=0.45\textwidth]{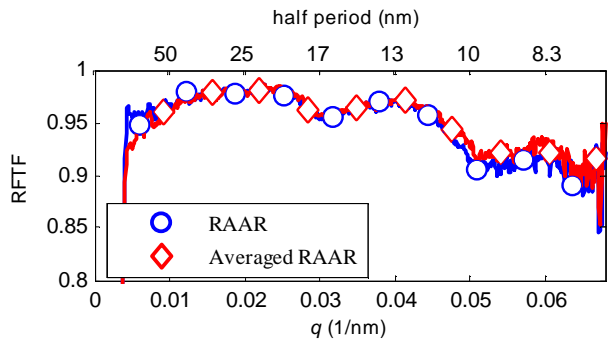}}
\caption{The R-Factor Transfer Function (RFTF) of a reconstructed image, showing excellent agreement with the measured diffraction intensities.}
\label{fig5} 
\end{figure}

An algorithm that always recovers the same phases does not necessarily 
recover the correct ones. Another requirement is that the recovered image is 
constrained in the region called support: $g\left( x \right) = 0,\;x \notin 
S$. If this condition is satisfied the Fourier modulus condition ($\vert 
G\vert \, = I^{1 / 2})$ is unlikely to be satisfied in the presence of 
noise. We can quantify deviations from the measured values by an $R$-factor 
(similar to that used in crystallography \cite{giacovazzo:2002}) by
\begin{equation}
\label{eq8}
\sigma _{RF}^{2} \left( {\rm {\bf q}} \right) = \frac{\left| {\left| 
{G\left( {\rm {\bf q}} \right)} \right| - \sqrt I \left( {\rm {\bf q}} 
\right)} \right|^2}{\left| {\sqrt I \left( {\rm {\bf q}} \right)} 
\right|^2}
\end{equation}
and its related transfer function $RFTF\left( {\rm {\bf q}} \right) = 
\left[ {1 + \sigma _{RF}^{2} \left( {\rm {\bf q}} \right)} \right]^{ - 1 / 
2}$, which is plotted in Fig. \ref{fig5} for a reconstructed image.

\section{Conclusion}
We have performed a characterization of high-resolution imaging of an 
isolated 3D object by \textit{ab initio} phase retrieval of the coherent X-ray diffraction, 
and examined metrics to allow the quality of image reconstructions to be 
assessed.

The phase retrieval process does not produce unique images, in that varying 
low-order phase modes arise, akin to aberrations in an imaging system. Other 
than the tilt terms, the low-order phase aberrations discussed here will be 
reduced in case of a real object (for which only antisymmetric terms are 
allowed) and will not be present when a real-space positivity constraint can 
be imposed, since defocusing or otherwise aberrating an image causes it to 
be complex. However, in the case of samples consisting of more than one 
material (such as biological samples) the object cannot be considered 
positive and we must reduce the effects of aberrations. We have proposed two 
methods of overcoming limitations of computer reconstruction: in order to 
improve the stability of the reconstructions we average several 
reconstructed images and use the result to feed a new round of phase 
retrieval. From an experimental point of view, the use of a reference point, 
or other well-defined object, should enable us to greatly reduce low order 
phase aberrations. 

\acknowledgments

Coccolith samples were provided by J. Young from the Natural History
Museum, London. This work was performed under the auspices of the
U.S. Department of Energy by University of California, Lawrence
Livermore National Laboratory under Contract W-7405-Eng-48 and the
Director, Office of Energy Research, Office of Basics Energy Sciences,
Materials Sciences Division of the U. S.  Department of Energy, under
Contract No. DE-AC03-76SF00098. This work has been supported by
funding from the National Science Foundation. The Center for
Biophotonics, an NSF Science and Technology Center, is managed by the
University of California, Davis, under Cooperative Agreement No. PHY
0120999.

\end{document}